\documentclass[atoms,article,accept,moreauthors,pdftex,12pt,a4paper]{mdpi}
\lastpage{\pageref*{LastPage}}
\doinum{10.3390/atoms3020182}
\pubvolume{3}
\pubyear{2015}
\history{Received: 7 April 2015  / Accepted: 18 May 2015  / Published: 26 May 2015}

\usepackage{graphicx}
\usepackage{color}
\usepackage{verbatim}
\usepackage[normalem]{ulem}
\usepackage{soul}
\usepackage{upgreek}

\def\be{\begin{equation}}
\def\ee{\end{equation}}
\def\ba{\begin{eqnarray}}
\def\ea{\end{eqnarray}}
\Title{Photon-Induced Spin-Orbit Coupling in Ultracold Atoms inside Optical Cavity}

\Author{Lin Dong $^{1}$, Chuanzhou Zhu $^{1}$ and Han Pu $^{1,2,}$*}

\address{%
$^{1}$ Department of Physics and Astronomy, Rice Quantum Institute, Rice University, Houston, \linebreak TX 77251-1892, USA; {E-Mails}: phydonglin@gmail.com (L.D.); ricezhuchzh@gmail.com (C.Z.) \\
$^{2}$ Center for Cold Atom Physics, Chinese Academy of Sciences, Wuhan 430071, China
\vspace{-12pt}}

\corres{E-Mail: hpu@rice.edu.}

\abstract{We consider an atom inside a ring cavity, where a plane-wave cavity field together with an external coherent laser beam induces a two-photon Raman transition between two hyperfine ground states of the atom. This cavity-assisted Raman transition induces effective coupling between atom's internal degrees of freedom and its center-of-mass motion. In~the meantime, atomic dynamics exerts a back-action to cavity photons. We investigate the properties of this system by adopting a mean-field and a full quantum approach, and show that the interplay between the atomic dynamics and the cavity field gives rise to intriguing nonlinear phenomena. }

\academiceditor{{Jonathan Goldwin  and Duncan O'Dell}}

\begin{document}

\vspace{-12pt}
\section{Introduction}

When Jaynes and Cummings first studied the time evolution of a two-level atom in an electromagnetic field in a {{ fully}} quantized way in 1960s \cite{JCM}, experimental realization of this ideal theoretical model was out of reach. It was made possible only with the advent of one-atom masers in late 1980s, by Rempe, Walther and Klein \cite{exp1987}, who experimentally studied the interaction of a single atom and a single resonant mode of electromagnetic field in a cavity. Jaynes-Cummings model (or J-C Model) serves to bridge our understanding of the relationship between quantum theory of radiation and semi-classical theory of atom-light interaction, and has become one of the most important models in quantum optics and cavity electrodynamics (CQED).
In the experiment of Reference~\cite{exp1987}, a beam of atoms was used and the atom-light interaction was studied during the transient time when the atoms pass through the cavity. The intensity of the atomic beam is sufficiently low such that at any given time, no more than one atom is inside the cavity.
The field of CQED was further advanced by putting a single trapped cold atom \cite{cavity0}, and more recently a condensate of ultracold atoms \cite{cavity1, cavity2, cavity3,Esslinger2010}, inside an optical cavity. In an optical ring cavity, through coherent controlling the dark and bright states, the vacuum Rabi splitting and optical bistability of cavity multi-wave-mixing process has been studied \cite{mixing1}. In addition, bright-state polaritons of four-wave mixing and six-wave mixing signals can be parametrically amplified \cite{mixing2}. In the J-C model, and other related CQED systems, the focus is the interaction and mutual influence between the cavity mode and the atomic internal degrees of freedom. The external degrees of freedom of the atom, \emph{i.e.}, its center-of-mass (COM) motion, is generally neglected. Unlike ``hot'' atoms, however, cold atoms' COM motion in general can no longer be neglected in this ``atom + cavity'' system, as the COM momentum of a cold atom will be significantly affected by photon recoil from emission and absorption of even a single photon. Therefore in a more complete description of the cavity system, one needs to take into account the interplay among the cavity photons, and both the internal and external atomic degrees of freedom. Furthermore, when more than one atom is inside the cavity, one should also account for the cavity photon mediated long-range interaction between atoms. All of these make the ``cold atoms + cavity'' system extremely rich and interesting, and truly represent a new frontier in both CQED and cold atom research.

As the same photon affects both the internal states (via inducing a transition between different states of the atom) and the external COM motion (via photon recoil) of the atom, it naturally induces a coupling between the two atomic degrees of freedom. Such spin-orbit coupling (SOC) in cold atoms has been realized in both bosonic \cite{soc1, soc2} and fermionic systems \cite{soc3, soc4}, and has attracted tremendous attention in recent years \cite{socVictor}. In practice, to avoid spontaneous emission, SOC is induced between two hyperfine ground states of an atom via a pair of Raman laser beams. Due to its non-Abelian nature, SOC not only significantly affects the physics of a single atom, but, perhaps more importantly, also profoundly changes the properties of a many-body system. It is an essential ingredient underlying such diverse phenomena as topological superconductors/insulators, Majorana and Weyl fermions, spin-Hall effects, \emph{etc.} \cite{TI, MF, WF, SHE1, SHE2}.

So far, all the experimental realization of photon-induced SOC in cold atoms employs classical laser fields to generate the Raman transition. Here the parameters of SOC are determined once the laser parameters (e.g., intensity, frequency, \emph{etc.}) are fixed. In a recent work \cite{cavitySOC}, we propose to replace one of the Raman beams by a plane-wave cavity mode supported by a ring cavity. Around the same time, Mivehvar and Feder considered a situation where both Raman beams are provided by a cavity (\cite{Feder}). In~this scheme, as the back-action from the atoms affects the cavity photons, the SOC thus induced becomes nonlinear and dynamical. By employing a simple mean-field approach, we demonstrated \cite{cavitySOC} that such a system indeed possesses interesting nonlinear properties even with just one single atom inside the cavity. For example, the cavity-assisted SOC dramatically modifies the atomic dispersion relation, in particular, with the emergence of loop structures under certain circumstances. Several other groups have recently investigated the cavity-assisted SOC in many-body atomic systems \cite{csoc1,csoc2,csoc3,csoc4}.

In the present work, we first briefly review our previous proposal \cite{cavitySOC}, and then theoretically explore the full quantum mechanical treatment beyond the mean-field formalism, and finally investigate the correspondence between the quantum and the mean-field treatment. The quantum treatment is carried out by solving the Master equation for the total density operator, from which we can derive various quantities of interest, \emph{i.e.}, the cavity photon statistics, the degree of entanglement between the atom and cavity field, \emph{etc.} These two different approaches provide a deeper understanding to this intriguing~system.

The article is organized as the following: After briefly reviewing key ideas of our previous work and the mean-field approach in Section \ref{meanfield}, we develop the full quantum mechanical formalism to the physical system of interest in Section \ref{master} and discuss about the intimate correspondence between the two in Section~\ref{relation}, and finally conclude in Section \ref{conclusion}.

\section{Model Setup and Mean-Field Formalism} \label{meanfield}

\begin{figure}[H]
\centering
\includegraphics[width=0.5\textwidth]{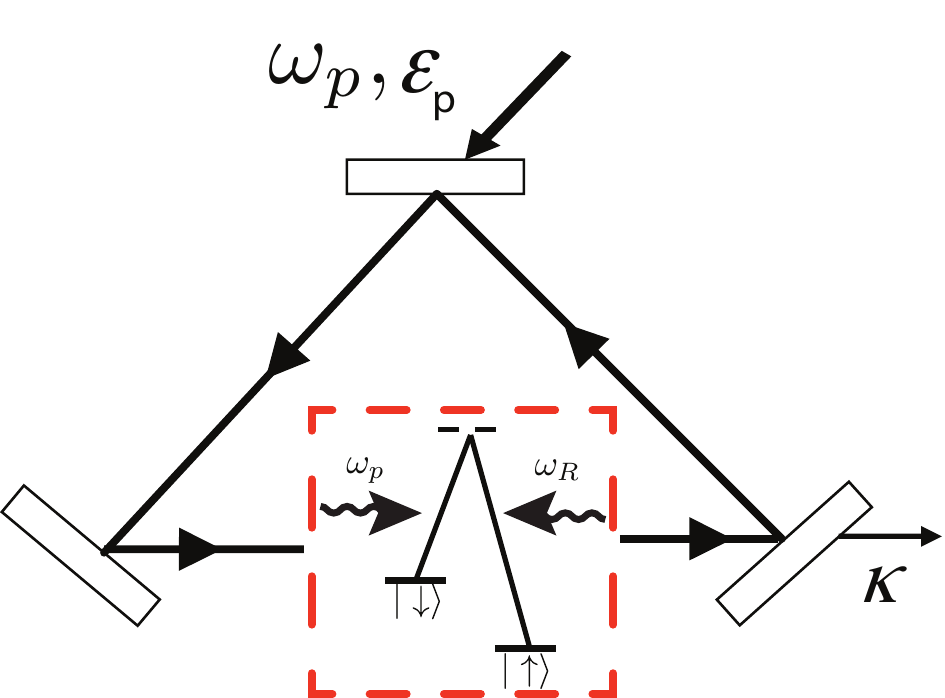} \vspace{12pt}
\caption{Schematic diagram of the cavity-assisted spin-orbit coupled system. Inside the red dashed box, we show the level diagram of the atom and the light field configuration where $\upomega_p$ represents the cavity field and $\upomega_R$ is an external coherent laser beam. $\upvarepsilon_p$ and $\kappa$ are the cavity pumping rate and decay rate, respectively.}\label{schematic}
\end{figure}

As shown schematically in Figure~\ref{schematic}, we consider a single atom being confined by a single-mode unidirectional ring cavity, whose cavity mode together with an additional coherent laser beam form a pair of Raman beams that induces transition between two hyperfine atomic ground states denoted as $|\uparrow\rangle$ and $|\downarrow\rangle$, while transferring recoil momentum $\pm 2\hbar q_r\hat{z}$ to the atom. The ring cavity has a resonant frequency of $\upomega_c$, decay rate $\kappa$, and is pumped by an external laser field with frequency $\upomega_p$ and pumping rate $\upvarepsilon_p$.
The Hamiltonian under the rotating wave approximation can be written as \cite{cavitySOC},
\ba
 \mathcal{H}_{\rm eff}& = & \sum_{\upsigma=\uparrow,\downarrow} \int d{z}\left[ \hat{\uppsi}^\dagger_\upsigma({z})\left(\frac{ { k}^2+ 2\upalpha_\upsigma q_r {k}}{2m}+\upalpha_\upsigma \updelta\right)\hat{\uppsi}_\upsigma(z)\right]+  \frac{\Omega}{2}\int d{z}\left[\hat{\uppsi}_{\uparrow}^{\dagger}({z})\hat{\uppsi}_{\downarrow}({z}) \hat{c}+ h.c. \right]\nonumber \\
 & + & i\upvarepsilon_{p}(\hat{c}^{\dagger}-\hat{c})-\updelta_c \hat{c}^{\dagger}\hat{c} \, \label{effH}
 \ea
where, for simplicity, we only consider the atomic COM motion along the $z$-axis, which is the direction of the photon recoil. In Hamiltonian Equation (\ref{effH}), $k$ is the atomic COM quasi-momentum (we take $\hbar =1$) along the $z$-axis, $\hat{c}$ is the cavity annihilation operator, $\hat{\uppsi}_\upsigma(z)$ ($\upsigma = \uparrow$, $\downarrow$) is the atomic operator, $\upalpha_{\uparrow, \downarrow}=\pm 1$, $2\updelta$ represents the two-photon Raman detuning, $\updelta_c=\upomega_p-\upomega_c$ is the cavity-pump detuning, and $\Omega$ denotes the atom-cavity coupling strength (or more specifically, the Raman coupling strength per cavity photon). Finally we will treat the cavity decay phenomenologically, which amounts to adding a non-Hermitian term $-i\kappa \hat{c}^{\dagger}\hat{c}$ in the above effective Hamiltonian. Note that for a quasi-momentum $k$, the real momentum for an atom is spin-dependent: it is $k+q_r$ for $|\uparrow \rangle$ state and $k-q_r$ for $|\downarrow \rangle$ state.

From Hamiltonian~(\ref{effH}), one can easily obtain the Heisenberg equations of motion for both atomic and cavity fields. In this work, we only consider a single atom inside the cavity. Hence we rewrite the atomic field operators as first quantization wave functions $ \hat{\uppsi}_\upsigma ({z}) \rightarrow  \uppsi_\upsigma (z)$. Assuming spatial homogeneity, we further take the plane-wave wave function for the atomic modes $\uppsi_\upsigma({z})=e^{i{k} {z}}\upvarphi_\upsigma$ with the normalization condition $|\upvarphi_\uparrow|^2+|\upvarphi_\downarrow|^2=1$. To proceed further, we adopt the mean-field approximation by replacing the photon field operators by its respective expectation value: $\hat{c} \rightarrow c\equiv \langle \hat{c} \rangle$. The validity of this mean-field approximation lies on two assumptions: (1) the cavity field can be described as a coherent state; (2) the atomic field and the cavity field have negligible entanglement. We shall come back to these two assumptions later when we compare the mean-field results with results obtained from the beyond-mean-field quantum Master equation.

Within the mean-field formalism, the steady-state solution for the photon field is obtained by taking the time derivative of the photon field to be zero, from which we obtain:
\begin{equation}
c = \frac{\upvarepsilon_{p}-i\frac{\Omega}{2} \upvarphi_\downarrow^\ast\upvarphi_\uparrow}{\kappa-i\updelta_{c}}
\label{c}
\end{equation}
Inserting this into the equations for atomic fields, we obtain the coupled nonlinear time-dependent equations for the two spin components as,
\ba
i\dot{\upvarphi}_{\uparrow}&\!\! =\!\! & \left(\frac{{k}^{2}}{2m}+q_{r}k+\updelta\right) \upvarphi_{\uparrow}+ \frac{\Omega_{\rm eff}}{2}\, \upvarphi_{\downarrow} \,
\label{EOMphi1}\\
i\dot{\upvarphi}_{\downarrow} & \!\!= \!\!& \left(\frac{{k}^{2}}{2m}-q_{r}k-\updelta\right)\upvarphi_{\downarrow}+ \frac{\Omega^*_{\rm eff}}{2}\,\upvarphi_{\uparrow}\,\label{EOMphi2}
\ea
where $\Omega_\text{eff} \equiv {\Omega} c ={\Omega} \frac{\upvarepsilon_{p}-i\frac{\Omega}{2} \upvarphi_\downarrow^\ast\upvarphi_\uparrow}{\kappa-i\updelta_{c}} $ is the effective Raman coupling strength between two atomic states. The fact that $\Omega_{\rm eff}$ depends on the atomic field itself is a manifestation of the non-linearity arising from the atomic back-action to the cavity field.

For a given atomic quasi-momentum ${k}$, we define eigenstate and eigenenergy as the solution of the time-independent version of Equations~(\ref{EOMphi1}) and (\ref{EOMphi2}), by replacing $i(\partial/\partial t)$ with $\upepsilon({k})$. After some lengthy but straightforward algebra, we find that $\upepsilon({ k})$ obeys a quartic equation in the form of
\begin{equation}
4\upepsilon^4+B\upepsilon^3+C\upepsilon^2+D\upepsilon+E=0 \,
\label{generalquarticEq}
\end{equation}
The derivation of the above equation and the explicit expressions of the $k$-dependent coefficients $B$, $C$, $D$ and $E$ can be found in our previous work \cite{cavitySOC}.

\begin{figure}[H]
\centering
\includegraphics[width=0.9\textwidth]{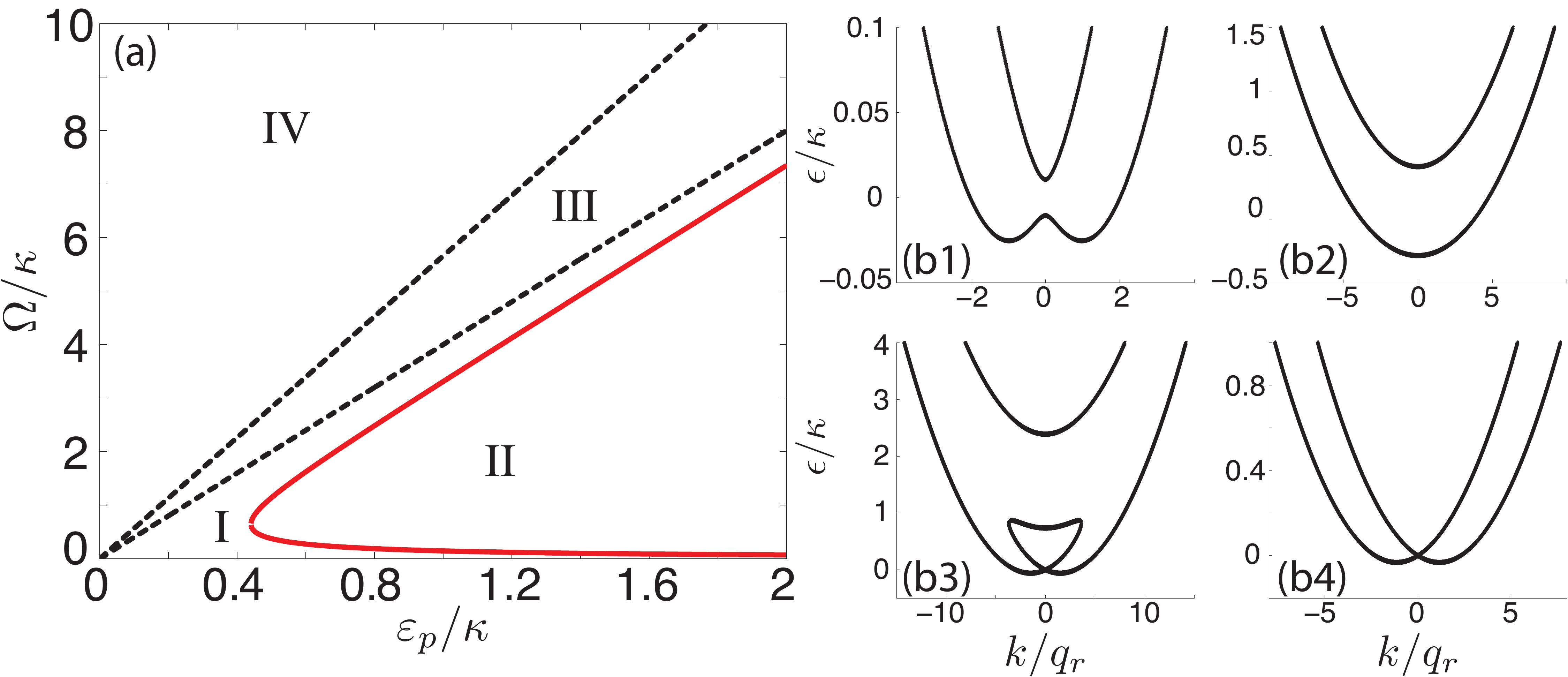}  \vspace{12pt}
\caption{Single particle eigen-energy spectrum ``phase diagram''.  The dispersion curve is generally catagorized by four regions, represented from I to IV in (\textbf{a}). From (\textbf{b1}) to (\textbf{b4}), we fix $\upvarepsilon_p=\kappa$. In region I, the dispersion has double minima as shown in (\textbf{b1}) with $\Omega=0.03$ $\kappa$; region II is enclosed by the red solid curve in (\textbf{a}) and we show the typical point in (\textbf{b2})  ($\Omega=\kappa$) where only single minimum exisits in the lower helicity branch; region III is enclosed by the black dashed lines in (\textbf{a}) and it is a region where loop structure emerges, as in (\textbf{b3}) with $\Omega=5$ $\kappa$; finally, in region IV we recover the double minimum dispersion although it's different from region I by closing the gap at $k=0$, as in (\textbf{b4}) with $\Omega=8$ $\kappa$. Throughout the paper we fix $\updelta_c=\kappa$ and $\updelta=0$, and adopt a dimensionless unit system with $\hbar=m=\kappa=1$. A typical value for $\kappa$ is $2\pi \times 1 \text{ MHz}$, and we choose $q_r = 0.22$ in our dimensionless units (based on a realistic experimental parameter estimate). }\label{fig1}
\end{figure}

We can gain some insights about the general structure of the dispersion relation $\upepsilon(k)$, e.g., the degeneracy condition and the appearance and disappearance of the loop. In the case of vanishing two-photon detuning (\emph{i.e.}, $\updelta=0$),
simple analysis shows that there should be a total of four regimes, as shown in Figure~\ref{fig1}a. In region I, the two dispersion branches are gapped, and the lower branch has a double degenerate minima, as shown in Figure~\ref{fig1}(b1). This dispersion curve structure is very similar to the case when both Raman beams are provided by classical coherent laser fields (we shall refer to this as the ``classical case'') and the Raman coupling strength is small. In region II, as shown in Figure~\ref{fig1}(b2), the two dispersion branches are still gapped, but the lower branch has a single minimum. This is similar to the classical case with a large Raman coupling strength. Regions III and IV do not have analogs in the classical case. Region III features a loop structure, as shown in Figure~\ref{fig1}(b3), whereas in Region IV, the loop dissolves but the two dispersion branches becomes gapless at $k=0$, as shown in Figure~\ref{fig1}(b4). In~the looped region, the quartic Equation (\ref{generalquarticEq}) yields four real roots. It can be shown \cite{cavitySOC} that this requires the coupling strength $\Omega$ to satisfy
\[\Omega_c^{(1)}\equiv4\upvarepsilon_p\leq \Omega \leq  4\upvarepsilon_p \sqrt{1+(\updelta_c/\kappa)^2}\equiv\Omega_c^{(2)} \,\]
The two dashed lines in Figure~\ref{fig1}a represent the two critical values $\Omega_c^{(1)}$ and $\Omega_c^{(2)}$, respectively.

\begin{figure}[H]
\centering
\includegraphics[width=0.9\textwidth]{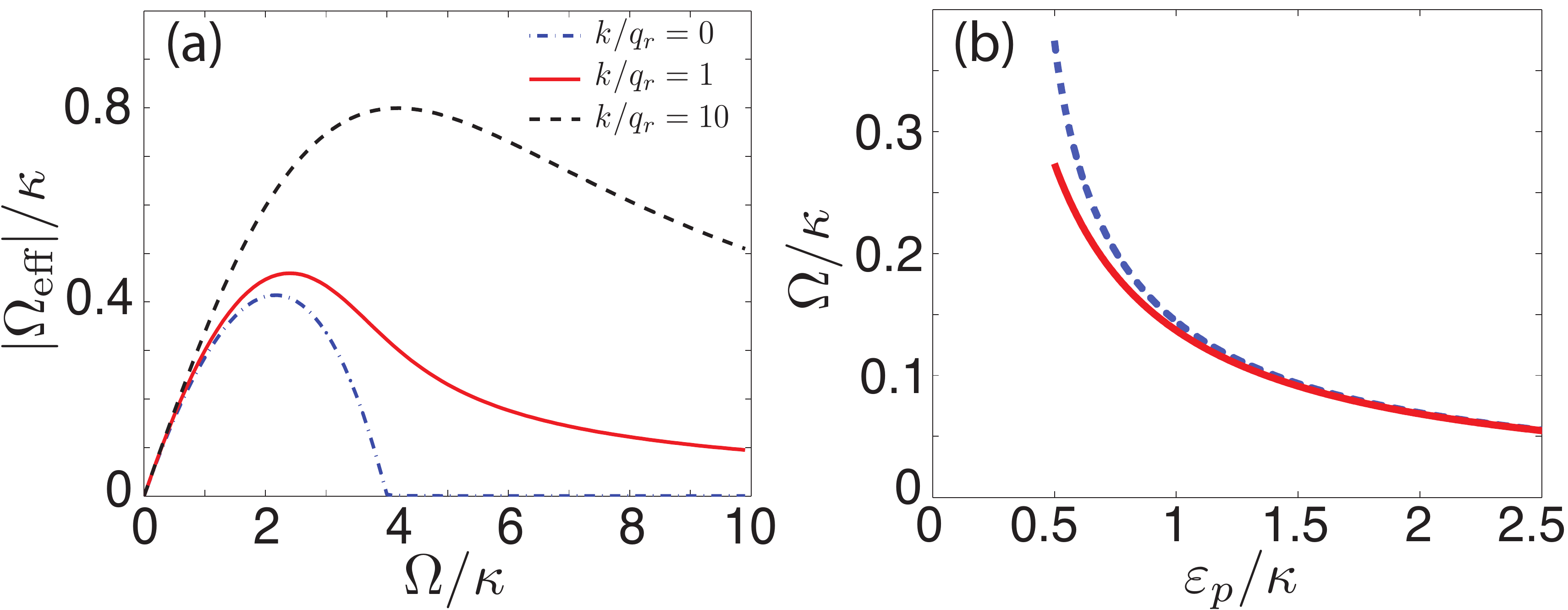} \vspace{12pt}
\caption{(\textbf{a}) Effective Raman coupling $|\Omega_\text{eff}|$ is plotted as a function of atom-photon coupling strength $\Omega$ for different $k$ values, $0$,\,$q_r$,\,$10$ $q_r$ for blue dash-dot, red solid and black dashed  lines. We observe that $|\Omega_\text{eff}|$ does not monotonically  increases with $\Omega$ but rather peaks at an intermediate value, then approaches zero in the large $\Omega$ limit. Figure (\textbf{b}) shows a comparison between critical boundary of region I and II (red solid curve) and the analytical result (blue dashed line) given in Equation~(\ref{ome}). At large $\upvarepsilon_p$ limit, the two results match asymptotically well.  }\label{fig2}
\end{figure}

It is instructive to examine how the effective Raman coupling strength $\Omega_{\rm eff}$ behaves as a function of $\Omega$. In Figure~\ref{fig2}a, we plot $|\Omega_\text{eff}|$ for the lowest dispersion branch as a function of $\Omega$ for different $k$ values. Note that $|\Omega_{\rm eff}|^2 = \Omega^2 n_{\rm photon}$ where $n_{\rm photon}=|c|^2$ is the steady-state cavity photon number. A few remarks can be made based on this plot. First, the fact that $\Omega_{\rm eff}$ is different for different $k$ clearly shows the influence of the atomic COM motion on both the atomic internal dynamics and the cavity photon number. Second, $|\Omega_{\rm eff}|$ is not a monotonous function of $\Omega$. For given $k$, $|\Omega_{\rm eff}|$ increases with $\Omega$ linearly for small $\Omega$. This can be intuitively understood as follows. At such weak atom-photon coupling, the back-action from the atom to the cavity photon is negligible. The number of cavity photons $n_{\rm photon}$ is roughly given by $n_{\rm photon} \approx n_0=\left|\frac{\upvarepsilon_{p}}{\kappa-i\updelta_{c}}\right|^2= \frac{\upvarepsilon_p^2}{\kappa^2+ \updelta_c^2}$, where $n_0$ is the number of cavity photons when the atom is absent. As a result, we have $|\Omega_{\rm eff}| \approx \Omega \sqrt{n_0}$ which is independent of the atomic quasi-momentum $k$. On the other limit, when $\Omega$ is very large, the strong atom-cavity coupling strength significantly detunes the cavity away from resonance and the cavity photon number $n_{\rm photon}$, and hence $|\Omega_{\rm eff}|$, decreases as a function of $\Omega$. Such a non-monotonous behavior of $\Omega_{\rm eff}$ is a unique feature of the cavity system and a direct manifestation of the non-linearity of the system arising from the back-action of the atom on the cavity photon.

From the above analysis, it should also become clear that when the effect of the back-action is weak (which occurs when $\Omega$ is small and/or $\upvarepsilon_p$ is large), we should recover the properties of the classical case. In particular, in the classical case, the lower dispersion branch change from two degenerate minima to a single minimum when the Raman coupling strength exceeds a critical value. Using our notation, this occurs when $|\Omega_{\rm eff}|$ exceeds the critical value $4E_r$ where $E_r=q_r^2/(2m)$ is the photon recoil energy. For~weak atom-cavity coupling, $|\Omega_{\rm eff}| = \Omega \sqrt{n_0}={\Omega} \frac{\upvarepsilon_p}{\sqrt{\kappa^2+ \updelta_c^2}}$. Hence the critical value of $\Omega$ is given by
\begin{equation}
\Omega=4E_r\frac{\sqrt{\kappa^2+\updelta_c^2}}{\upvarepsilon_p}\,
\label{ome}
\end{equation}

In Figure~\ref{fig2}b, we plot this critical value (blue dashed line) as a function of cavity pump rate $\upvarepsilon_p$ and compare it with the numerically determined lower boundary (red solid line) between region I and II of Figure~\ref{fig1}a. The two curves overlap with each other when $\upvarepsilon_p$ increases. Therefore, as we have expected, in the limit of weak atom-cavity coupling and strong cavity pumping, we fully recover the classical case where the SOC is induced by two classical laser beams.

\section{Master Equation Approach: Full Quantum Mechanical Treatment } \label{master}

The above discussion is based on the mean-field approach where the cavity field is replaced by a $c$-number that represents the photon amplitude. This mean-field treatment relies on two implicit assumptions: (1) the atom-photon correlation is negligible, and (2) the photon field can be well approximated by a coherent state. In order to examine the validity of these assumption, and hence the validity of the mean-field approximation, we now turn to a full quantum treatment based on the Master~equation:
\be
\dot{\rho} = \frac{1}{i\hbar}[{\cal H}_{\text{eff}},\rho]+\mathcal{L}[\rho] \, \label{masterEq}
\ee
Here $\rho$ is the total density operator of the coupled atom-cavity system, the effective Hamiltonian ${\cal H}_{\text{eff}}$ is the same as in Equation~(\ref{effH}). The dissipation arising from cavity decay is modeled by the Liouvillean term in the
standard form of Lindblad super-operator \cite{L1, L2},
\be
\mathcal{L}[\rho] = \kappa (2c\rho \hat{c}^\dagger-\hat{c}^\dagger \hat{c}\rho-\rho \hat{c}^\dagger \hat{c})\,\label{Lindblad}
\ee
Again, due to spatial homogeneity, we decouple momentum eigenstates by taking the plane-wave ansatz for the atomic modes as $\hat{\uppsi}_\upsigma({z})=e^{i{k}{z}}\hat{\uppsi}_\upsigma$. As there is no coupling between atomic operators with different $k$, we can work in the subspace for a fixed value of $k$. Here we explicitly write the commutator in Equation~(\ref{masterEq}), for a given atomic quasi-momentum $k$, as,
\ba
[{\cal H}_{\text{eff}}({k}),\rho] & = & \left(\frac{{ k}^{2}}{2m}+\frac{q_{r}k}{m}+\updelta\right)\left(\hat{\uppsi}_{\uparrow}^{\dagger} \hat{\uppsi}_{\uparrow}\rho-\rho \hat{\uppsi}_{\uparrow}^{\dagger} \hat{\uppsi}_{\uparrow}\right)+\left(\frac{{k}^{2}}{2m}-\frac{q_{r}k}{m}-\updelta\right) \left(\hat{\uppsi}_{\downarrow}^{\dagger} \hat{\uppsi}_{\downarrow}\rho-\rho \hat{\uppsi}_{\downarrow}^{\dagger} \hat{\uppsi}_{\downarrow}\right)\nonumber\\
 &  & + \frac{\Omega}{2}\left(\hat{\uppsi}_{\uparrow}^{\dagger} \hat{\uppsi}_{\downarrow} \hat{c} \rho+\hat{c}^{\dagger} \hat{\uppsi}_{\downarrow}^{\dagger} \hat{\uppsi}_{\uparrow}\rho-\rho \hat{\uppsi}_{\uparrow}^{\dagger} \hat{\uppsi}_{\downarrow}c-\rho \hat{c}^{\dagger} \hat{\uppsi}_{\downarrow}^{\dagger} \hat{\uppsi}_{\uparrow}\right)\nonumber\\
&& + i\upvarepsilon_{p}\left(\hat{c}^{\dagger}\rho-\hat{c}\rho-\rho \hat{c}^{\dagger}+\rho \hat{c}\right)-\updelta_{c}\left(\hat{c}^{\dagger}\hat{c}\rho-\rho \hat{c}^{\dagger}\hat{c}\right)\,
\ea
Note that if the photon recoil $q_r=0$, which occurs when the cavity mode and the external laser beams are co-propagating, the COM kinetic energy terms $k^2/(2m)$ can be gauged away after a simple gauge transformation. Our model is then reduced to the J-C model and the atomic COM motion does not play a role.
To solve the Master Equation (\ref{masterEq}), we choose our basis states as direct product states of photon Fock state $|n \rangle$ and atomic internal state $|\upsigma=\uparrow,\downarrow \rangle$: $|n;\upsigma\rangle \equiv |n \rangle \otimes |\upsigma \rangle$, 
where non-negative integer $n$ denotes cavity photon number. Our goal is to calculate the entire matrix elements of the density operator under this set of basis states, denoted by $\langle m;\upsigma|\rho|n;\upsigma'\rangle\equiv\rho_{mn}^{\upsigma\upsigma'}$. We found the governing equation for the matrix element can be written as,

\ba
\frac{d}{dt}\rho_{mn}^{\upsigma\upsigma'}
& = & -i\left(\frac{{k}^{2}}{2m}+\frac{q_{r}k}{m}+ \updelta\right)\left(\updelta_{\upsigma\uparrow}-\updelta_{\upsigma'\uparrow}\right)\rho_{mn}^{\upsigma\upsigma'}
-  i\left(\frac{{k}^{2}}{2m}-\frac{q_{r}k}{m}-\updelta\right)\left(\updelta_{\upsigma\downarrow}-\updelta_{\upsigma'\downarrow}\right)\rho_{mn}^{\upsigma\upsigma'}\nonumber\\
& + & \frac{\Omega}{2i}(\updelta_{\upsigma\uparrow}\sqrt{m+1}\rho_{m+1n}^{\bar{\upsigma}\upsigma'}
+\updelta_{\upsigma\downarrow}\sqrt{m}\rho_{m-1n}^{\bar{\upsigma}\upsigma'}
-\updelta_{\upsigma'\uparrow}\sqrt{n+1}\rho_{mn+1}^{\upsigma\bar{\upsigma'}}
-\updelta_{\upsigma'\downarrow}\sqrt{n}\rho_{mn-1}^{\upsigma\bar{\upsigma'}})\nonumber\\
& + & \upvarepsilon_{p}\left(\sqrt{m}\rho_{m-1n}^{\upsigma\upsigma'}-\sqrt{m+1}\rho_{m+1n}^{\upsigma\upsigma'}+\sqrt{n}\rho_{mn-1}^{\upsigma\upsigma'}-\sqrt{n+1}\rho_{mn+1}^{\upsigma\upsigma'}\right)\nonumber\\
& + & i\updelta_{c}\left(m-n\right)\rho_{mn}^{\upsigma\upsigma'}
+ \kappa\left(2\sqrt{m+1}\sqrt{n+1}\rho_{m+1n+1}^{\upsigma\upsigma'}-(m+n)\rho_{mn}^{\upsigma\upsigma'}\right)\label{EOMrho}
\ea
where $\bar{\upsigma}$ represents the flip-spin value, \emph{i.e.}, $\bar{\uparrow}=\downarrow$ and $\bar{\downarrow}=\uparrow$.

\begin{figure}[H]
\centering
\includegraphics[width=0.6\textwidth]{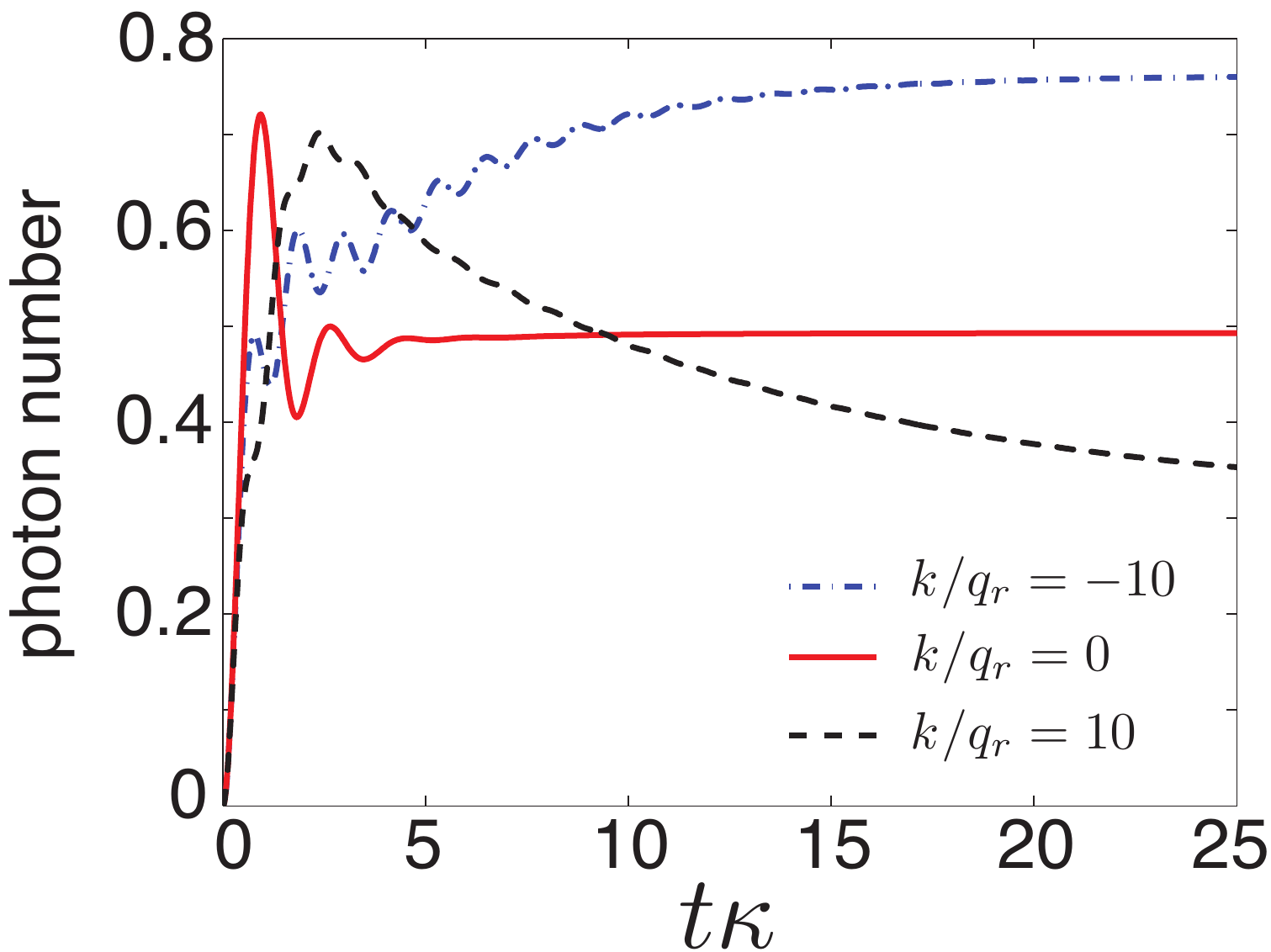}\vspace{6pt}
\caption{Time evolution of cavity photon number. The initial state is given by $|0;\uparrow\rangle$ and we consider the same parameter as in Figure~\ref{photon}a with $k/q_r=-10,\,0,\,10$. The steady-state values, obtained in the long time limit as shown here, correspond to red dashed lines at corresponding $k$ values in Figure~\ref{photon}a.}\label{evo}
\end{figure}

With Equation~(\ref{EOMrho}), we can study the dynamical evolution of the density operator $\rho$ for a given initial state. Obviously, we need to introduce a sufficiently large photon number cutoff. Once we obtain the density operator, all relevant physical quantities can be readily calculated. An example is given in Figure~\ref{evo}, where we show the time evolution of the cavity photon number $n = {\rm Tr}[\rho \hat{n}]= {\rm Tr}[\rho \hat{c}^\dag \hat{c}]$ for the initial state $|0; \uparrow \rangle$. The three different curves in Figure~\ref{evo} correspond to different atomic quasi-momentum~$k$.

\begin{figure}[H]
\centering
\includegraphics[width=0.9\textwidth]{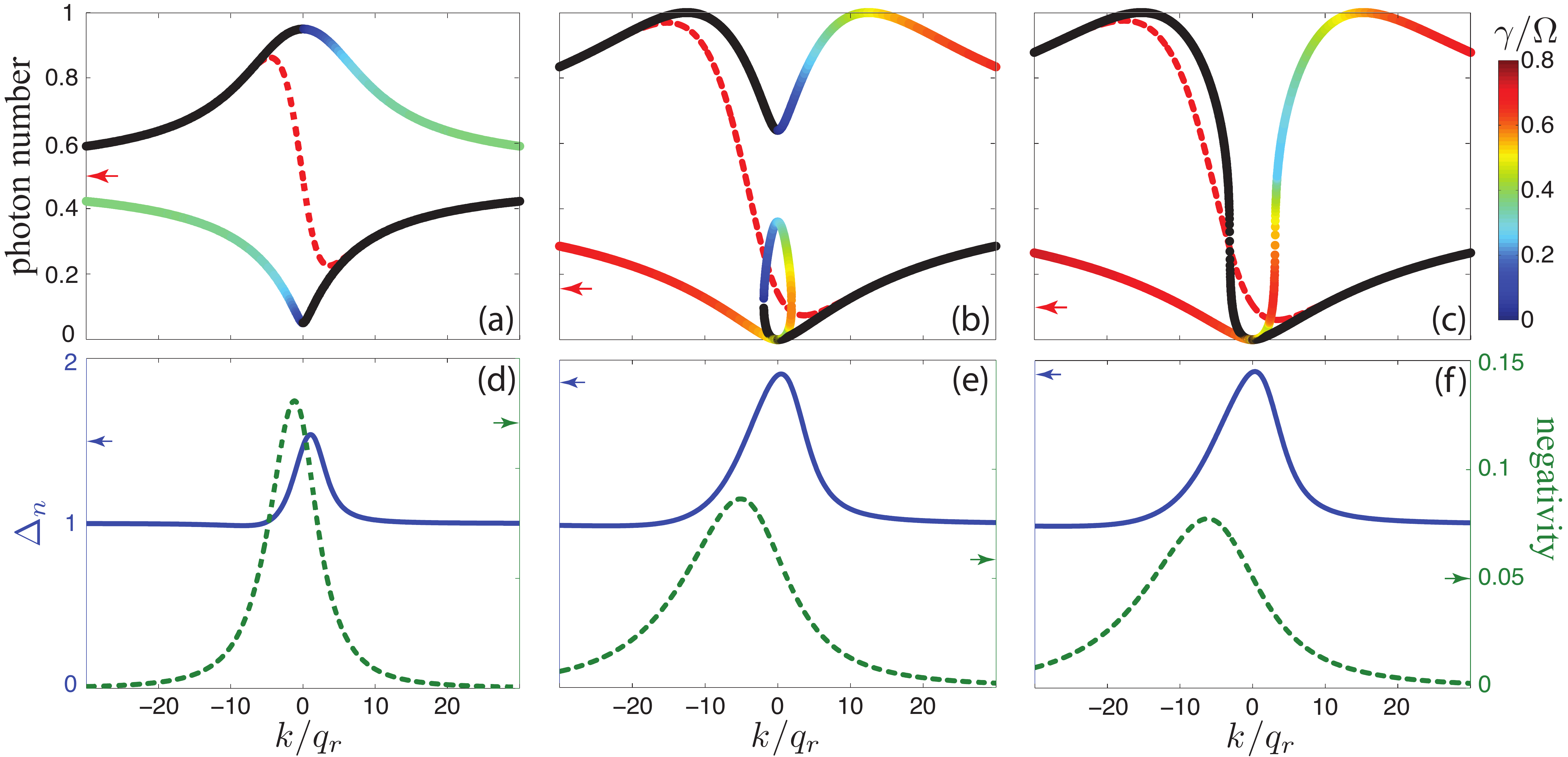}\vspace{6pt}
\caption{(\textbf{a})$\sim$(\textbf{c}) Photon number obtained from the mean-field approach (solid curves) and from quantum mechanical Master equation approach (red dashed curves). From (\textbf{a}) to (\textbf{c}), $\Omega = 3$ $\kappa$, $5.6$ $\kappa$, $6$ $\kappa$. The color on the solid curves represents the normalized decay rate $\gamma/\Omega$ of unstable mean-field states. The black color represents stable mean-field states. We~have used $\upvarepsilon_p=\kappa$, and other parameters are the same as in Figure~\ref{fig1}. (\textbf{d})$\sim$(\textbf{f}) Corresponding photon number fluctuation (blue solid curve) and negativity (green dashed line) obtained from the quantum approach. The parameters used here are the same in (\textbf{a})$\sim$(\textbf{c}), respectively. The~horizontal arrows indicate results from the J-C model by taking $q_r=0$.
}\label{photon}
\end{figure}

As evidenced in Figure~\ref{evo}, due to the presence of cavity decay, a steady state will eventually be reached. Let us now focus on the properties of the steady state. The steady-state density operator matrix elements can be obtained by equating the RHS of Equation~(\ref{EOMrho}) to zero. The red dashed lines in Figure~\ref{photon}a--c represent the steady-state photon number as functions of atomic quasi-momentum $k$. The horizontal arrows in the plot represent the cavity photon number by setting $q_r=0$, in which case our model reduces to the J-C model and all physical quantities become $k$-independent.

To have a better understanding of the photon statistics, we study the steady-state photon number fluctuation. Specifically, we calculate the normalized photon number fluctuation defined as
\[\Delta_n = \frac{\langle(\Delta n)^{2}\rangle}{\langle \hat{n}\rangle}=\frac{\langle \hat{n}^{2}\rangle-\langle \hat{n}\rangle^{2}}{\langle n\rangle} \,\]
where the expectation values of the operators are obtained with the help of the steady-state density operator. For a coherent cavity field, the photon fluctuation is Poissonian and we have $\Delta_n=1$.
The~solid curves in Figure~\ref{photon}d--f represent $\Delta_n$ (left vertical axis) as functions of $k$, and the horizontal arrows pointing to left give the values of $\Delta_n$ from the J-C model by setting $q_r=0$. For the parameters we have used, we note that the J-C model always predicts a super-Poissonian photon statistics, whereas our model gives super-Poissonian photon statistics only for small atomic quasi-momentum, but Poissonian statistics as $k/q_r \rightarrow \pm \infty$.

Last but not least, to characterize the correlation between the atom and the cavity field, we investigate the so-called negativity \cite{negativity} which measures the degree of entanglement for a mixed state system.
The~negativity is defined as $\mathcal{N}(\rho)=(||\rho^{T_A}||_1-1)/2$, where $\rho^{T_A}$ is the partial transpose of the density operator with respect to either the atom subsystem or the cavity subsystem, and $||\rho^{T_A}||_1$ denotes its trace norm with the definition $||\hat{A}||_1 \equiv {\rm Tr}[\sqrt{\hat{A}^\dag \hat{A}} ]$. A negativity of zero indicates that the two subsystems (the atom and the cavity, in our case) are not entangled, whereas a positive negativity means that finite degree of entanglement is present. The dashed curves in Figure~\ref{photon}d--f represent the negativity (right vertical axis) in the steady state as functions of $k$, and the horizontal arrows pointing to right give the values of the negativity from the J-C model by setting $q_r=0$. One can observe that for the chosen parameters, the J-C model always predicts a finite degree of entanglement between the atom and the cavity field. By~contrast, the degree of entanglement in our model weakens when $k/q_r \rightarrow \pm \infty$.

\section{Discussions} \label{relation}

In the previous two sections, we have presented two different methods for studying the system. We~are now in a position to discuss their connections.

In Figure~\ref{photon}a--c, in addition to the steady-state photon number obtained from the quantum treatment (dashed curves), we also plot the photon number $n_{\rm photon}=|c|^2$ obtained from the mean-field approach (solid curves), with $c$ given in Equation~(\ref{c}). In the quantum treatment, the steady-state density matrix is obtained by solving a set of linear equations. For a given $k$, the solution is unique. Hence we only get one steady-state photon number for a given atomic quasi-momentum. On the other hand, the mean-field treatment allows multiple steady-state solutions corresponding to different real roots of the quartic equation Equation~(\ref{generalquarticEq}). Hence a single $k$ value is associated with more than one steady-state photon number. However, due to the non-linearity intrinsic in the mean-field method, not all steady-states are dynamically stable. A straightforward stability analysis allows us to quantify the dynamical stability of the mean-field states, as we did in Reference~\cite{cavitySOC}. The stability information of the mean-field states are encoded as the color value in the solid curves. Those stable states are represented by black color, while any color other than black indicates an unstable state, and the color value represents the decay rate (see the colorbar) of the corresponding state. From Figure~\ref{photon}a--c, we clearly see that at large atomic quasi-momentum $k/q_r \rightarrow  \pm \infty$, the Master equation result overlaps with the stable mean-field branch; while at small $|k|$, the quantum result deviates significantly away from the mean-field solution.

The agreement for large $|k|$ and the discrepancy at small $|k|$ are both consistent with the results of the negativity and photon number fluctuations as presented in Figure~\ref{photon}d--f: At large $|k|$, the negativity is small (\emph{i.e.}, atom-cavity entanglement is weak) and the photon number fluctuation tends to Poissonian (\emph{i.e.}, the photon field is well approximated by a coherent state), this is exactly the regime where we expect the mean-field approximation is valid. By contrast, for small $|k|$, the quantum calculation indicates that there is non-negligible entanglement between the atom and the cavity field, and the cavity field itself cannot be assumed as a coherent state. Hence the mean-field assumption is no longer valid.

The reason why mean-field approximation only works for large $|k|$ is actually rather simple. Consider a Raman transition process where the atom jumps from $|\uparrow \rangle$ to $|\downarrow \rangle$. The quasi-momentum $k$ does not change during this process, however the real momentum changes from $k+q_r$ to $k-q_r$. Therefore the effective two-photon Raman detuning is not just $2\updelta$, but $2\updelta +2q_rk/m$, where the additional term comes from the difference of the kinetic energies for different pseudo-spin state $|\upsigma\rangle$. In other words, the SOC renders the two-photon detuning momentum-dependent. In the examples we presented in this work, we have taken $\updelta=0$. Hence the Raman transition is only near-resonant for small $|k|$, and becomes far off-resonant for large $|k|$. Therefore, for large $|k|$, the atom-photon coupling, and hence the atomic back-action to cavity, are weak. This explains why the mean-field assumption becomes valid in this~regime.

\section{Conclusions} \label{conclusion}

In this work, we have studied spin-orbit coupled cold atoms inside a ring cavity system, employing both the mean-field theory and the full quantum mechanical Master equation approach. By treating both light and atom on equal footing and seeking the self-consistent solution in both approaches, we have found: (1) cavity-assisted SOC dramatically modifies atomic dispersion relation, (2) intriguing dynamical instabilities exist in the system, (3) atom's back-action onto cavity field also leads to non-trivial atom-photon coupling that is fundamentally different from either the system with classical-laser induced SOC in the absence of the cavity or the J-C model where the atomic COM motion is neglected (\emph{i.e.}, by taking $q_r=0$). We have also explored correspondence and discussed the connection between the mean-field and the quantum approaches. The two distinctively different approaches provide us with a deeper understanding and complementary insights into this system. We conclude that the synthesis of cavity QED and SOC is not a trivial combination and interesting new physics emerges in this setting.

\acknowledgments{Acknowledgments}
This work is supported by the US National Science Foundation and the Welch Foundation (Grant No.~C-1669).

\authorcontributions{Author Contributions}
 Han Pu conceived the idea of the project, Lin Dong and  Chuanzhou Zhu explored the theoretical and numerical aspects of the physics. All authors contributed to writing and revising the manuscript and participated in the discussions about this work.

\conflictofinterests{Conflicts of Interest}
The authors declare no conflict of interest.

\bibliographystyle{mdpi}
\makeatletter
\renewcommand\@biblabel[1]{#1. }
\makeatother

\end{document}